\documentclass[12pt,preprint]{aastex}

\shorttitle{Scaling in molecular clouds}
\shortauthors{Boldyrev, Nordlund \& Padoan}

\begin{document}
\input psfig.sty

\title{Scaling relations of
supersonic turbulence in star-forming molecular clouds}

\author{Stanislav Boldyrev}
\affil{{\em Institute for Theoretical Physics,
Santa Barbara, California 93106},\\ {\sf boldyrev@itp.ucsb.edu} }
\author{{\AA}ke Nordlund}
\affil{{\em  Copenhagen Astronomical Observatory and Theoretical
Astrophysics Center, DK-2100 Copenhagen, Denmark},\\ {\sf aake@astro.ku.dk}}
\author{Paolo Padoan}
\affil{{\em Harvard University, Department of Astronomy,
60 Garden street, Cambridge, MA 03138},\\ {\sf ppadoan@cfa.harvard.edu}}


\begin{abstract}
We present a direct numerical and analytical study
of driven supersonic MHD turbulence that is
believed to govern the dynamics of star-forming molecular clouds.
We describe statistical properties of the turbulence by measuring
the velocity difference
structure functions up to the fifth order. In particular,
the velocity power spectrum
in the inertial range is found to be close to~$E_k\sim k^{-1.74}$,
and the velocity difference scales
as~$\langle |\Delta u|\rangle \sim L^{0.42}$. The results agree well
with the Kolmogorov--Burgers analytical
model suggested for supersonic turbulence in~[astro-ph/0108300].
We then generalize the model to more realistic,
fractal structure of molecular clouds, and show that depending on
the fractal dimension of a given molecular cloud, the theoretical
value for the  velocity spectrum spans  the
interval~$[-1.74\cdots -1.89]$, while the corresponding window
for the velocity difference scaling exponent is~$[0.42\cdots 0.78]$.
\end{abstract}

\section{Introduction}
Observations and numerical simulations of gas motion in
molecular
clouds show rather complex distributions of velocity and
density fields, indicating that the motion is
turbulent. Although the nature and the
scales of the driving force of the turbulence can vary,
it has been established long ago that this turbulence is
highly supersonic, with Mach numbers varying from cloud to
cloud and reaching up to of the order of~$30$ at scales
$\sim 100$ pc
\citep{Larson3,Larson1,Larson2,Falgarone,Falgarone+92,%
Myers,Padoan-Nordlund2,PN,Ossenkopf,Ostriker}.

Star formation is, of course, ultimately due to
gravitational collapse of small Jeans-unstable cores.
However, as is seen in
high-resolution  numerical simulation,
the initial density fragmentation,
that leads to creation of such cores, may be mainly
due to strong supersonic turbulence, i.e. may be explained
to large extent without invoking the effects of
self-gravity \citep{Padoan-etal2}.
The process of star formation can therefore be divided into
two stages that may be approached separately.  In the first
stage, the supersonic, turbulent motion of the interstellar
gas develops shocks that interact with each other and with
the turbulent flow, which results in the emergence of
rather complicated density structures. This stage should
be described {\em statistically}, in terms of probability
distribution of density and velocity
fluctuations, or in terms of their moments; attempts
to understand the complicated picture of turbulence by studying
each particular structure can hardly be successful. On the other
hand, the second stage, dealing with collapsing cores, can be
approached on the grounds of classical dynamics, and
depends on the effects of gravity and of the specific
environment (e.g., pressure, temperature, magnetic fields).
These two stages are not independent, however, since the first one
sets the distribution of initial conditions for the second one, for
further discussion
see, e.g., \cite{Padoan-Nordlund2,Klein,Burkert,Klessen,Elmegreen}.
Even more, with some suitable definition of density clumps, the
first stage of density fragmentation leads to the distribution
of clumps over masses that already resembles
the observed initial mass distribution function of
stars \citep{PN,Padoan-etal}.

In this paper we address both numerically and analytically the
first, ``turbulent'' stage of star-formation. So far, there has
been no analytical theory predicting the statistical
properties of supersonic interstellar turbulence,
notwithstanding the fact that supersonic conditions have
been inferred from observations for more than 20 years.
In particular, the turbulence scaling relations,
referred to as the Larson's laws, concern the scaling of
velocity and density fluctuations with respect to the
size of the fluctuations \citep{Larson3,Larson1}.
These scalings seem to vary for different clouds, but most
observations suggest that the velocity difference scaling scatters
around~$\langle |\Delta u|^2\rangle^{1/2} \sim L^{0.4}$
\citep{Falgarone+92}. The corresponding velocity power
spectra are steeper than the Kolmogorov one~$E_k\sim k^{-5/3}$,
which would correspond to $\langle |\Delta u|^2\rangle^{1/2} \sim L^{1/3}$.
As for the density fluctuations, the scaling of the peak
density of fluctuations on scales $L$ is close to~$\rho (L)\sim L^{-1}$
\citep{Falgarone+92}.
However, these results should be
taken with a certain degree of precaution. The error bars of
available observations are rather large, and the systematic
errors are sometimes unknown. Moreover, there are fundamental
reasons that prevent one from restoring the three-dimensional
velocity
correlators from the measured two-dimensional
projections \citep{Ballesteros-Paredes}.

The high-resolution numerical results that recently became
available \citep[e.g.][]{PNJ97,Porter-etal2,MacLow,Porter-etal1,%
Stone,Padoan-Nordlund2,PN,Padoan-etal}
shed some light onto the physics of
supersonic turbulence. In this paper, we present numerical
simulations of supersonic, super-Alfv\'enic turbulence,
driven on large scales in such a way that the sonic Mach number $M$ is of
the order 10 and the Alfv\'enic Mach number $M_a$ is of the order 3.
An analysis of simulations with different
Mach numbers will be presented elsewhere \citep{Jimenez}. We propose
an analytical theory that explains the results of
our numerical findings, and  discuss its
applicability to molecular clouds. We are
interested  mostly in the velocity correlators, although
an application to the density statistics is presented as well.
The next section analyzes the results of the numerical simulations.
In particular, the velocity-difference structure functions are
constructed.  In section~\ref{theory} we show that the observed
features agree well with the recently proposed Kolmogorov--Burgers
theory of supersonic turbulence. Section~\ref{conclusions}
discusses application of the results to molecular clouds. Conclusions
are presented in section~\ref{conclusions1}.

\section{Scaling laws in numerical simulations of supersonic turbulence}
\label{numerics}

The numerical simulations were performed with~$250^3$ and~$500^3$
resolutions for MHD~turbulence with an isothermal equation of state,
using the same method and program as in \citet{PNJ97}, \citet{Padoan-Nordlund2},
\citet{Padoan-etal}, and \citet{Padoan-etal2}.
The turbulence was driven on large scales by a solenoidal external
force with $1\leq k\leq 2$, where $k=1$ corresponds to the size of
the periodic box.  The solenoidal character of the forcing
is not crucial for the velocity scaling in the inertial
region---this was checked by comparing with runs with mixed compressional
and solenoidal driving.   The solenoidal driving
was chosen to provide a better `boundary condition' for the inertial
interval in $k$~space, since it turns out that the compressional to
solenoidal ratio tends to become small in the inertial range.
The external force sustains the supersonic
gas motion ($M=10$) in the simulations.  The motions would otherwise
decay on a time scale of the order of a crossing time and become sub-sonic due to
the dissipation in shocks \citep{Stone,Padoan-Nordlund2}.  The
real forcing is probably due to a turbulent cascade from large scales,
driven by supernovae and superbubbles \citep{Korpi,Avillez},

The supersonic turbulence exhibits rather interesting
properties that we summarize as follows.  First of all, the spectra
of both the potential,~${\bf u}_c$, and the solenoidal,~${\bf u}_s$,
components of the velocity field are steeper than the Kolmogorov
spectrum~$k^{-5/3}$. These components are defined according
to~$\nabla \cdot {\bf u}_s=0$  and~$\nabla \times {\bf u}_c=0$.
In Fig.~\ref{spectra} we plot the solenoidal spectrum weighted by~$k^{1.74}$,
since the theory that we present below predicts for the velocity
spectrum of turbulence~$E_k \sim k^{-1.74}$.
Second, the divergence-free, solenoidal part of
the velocity field, ${\bf u}_s$, is generated quite
effectively by such turbulence, contrary to the two-dimensional case,
where turbulence without pressure  is mostly
potential. This effect of vorticity generation is analogous to
the magnetic dynamo effect existing in 3D and non-existing in
2D~turbulence \citep[also, the presence of a magnetic field can
help generate vorticity, as discussed by][]{Enrique}.
We find that
the compressible part accounts for only~10--20 percent of the
intensity of the velocity field, see Fig.~\ref{spectra}.
The ratio $\gamma=\langle u^2_c \rangle/\langle u^2_s\rangle $
can thus be chosen as a small parameter of the
turbulence.
Third,  the dissipative
structures look like
two-dimensional shocks rather than one-dimensional filaments or
vortices as in incompressible turbulence. To demonstrate this, we
plot in Fig.~\ref{shocks}  a randomly chosen two-dimensional
cross-section of the density distribution in a physical simulation domain.
The filaments seen on the picture correspond to the two-dimensional
shock structures.

To quantitatively
characterize the statistical properties of the turbulence we
have measured the so-called structure functions of the velocity field \citep{Frisch2}.
These functions are defined as:
\begin{eqnarray}
S_p(L)=\langle \vert u(x+L)-u(x) \vert^p \rangle \sim L^{\zeta(p)},
\end{eqnarray}
where~$u$ is the component of the velocity field perpendicular or
parallel to the vector~${\bf L}$. According to the chosen component,
the structure functions are called transversal or longitudinal,
respectively. In the inertial interval, the
structure functions obey scaling laws,
and the exponents~$\zeta(p)$ may be determined; it is usually
expected that both transversal and longitudinal functions have
the same scaling.
The power
spectrum of the velocity field is the Fourier transform of
the second-order structure function, and may be expressed
as~$E_k\sim k^{-1-\zeta(2)}$.

We performed measurements of the
structure functions up to~$p=5$.
Since the Reynolds number in
our simulations was not large enough
to observe good scaling behavior of the structure functions, the method
of Extended Self-Similarity (ESS) was applied. Namely, instead of
plotting the structure functions themselves, we plotted the ratios of
their logarithmic slopes. As was discovered by \citet{Benzi1} and \citet{Benzi2},
such ratios exhibit
rather good (and correct) scaling behavior, even in systems with
moderate Reynolds numbers. Then, if one knows from the theory that the
third order structure function scales with~$\zeta(3)=1$, or if
any other scaling exponent~$\zeta(n)$ is known from numerics
or observations
with good precision, one can obtain
the scalings of all the other structure functions.
To illustrate this procedure, in Fig.~\ref{structure} we
show the plots of the transversal
structure functions,~$S_1\dots S_5$, their logarithmic slopes,
and the ratios of the slopes vs that of $S_3$. We chose to deal with
the transversal rather than longitudinal
functions since the flow is {\em shear-dominated}, and
the transversal functions are therefore found to have a
better scaling behavior.
As illustrated by Fig.~\ref{structure},
the structure functions themselves do not exhibit good
scaling behavior, but their relative scaling is well
established in a rather large interval. As has been
suggested by \citet{Dubrulle},
the ratios of scaling exponents may be universal while the scaling
exponents themselves may be not. The
reason is
that the physical length~$L$ may not be a good scaling variable, and
the universal scaling of all structure functions
holds not with respect to~$L$, but with respect
to a certain function~$\xi(L)$ that depends on the Reynolds number
and other parameters of the system.

From Fig.~\ref{structure} we conclude that the observed properties of
the supersonic turbulence can
hardly be predicted in the framework of any known model
of strong turbulence.
Indeed, the Kolmogorov model, or its generalizations to MHD,
cannot be directly
applied, since the turbulence has very small pressure and
rather weak magnetic field.
The model of turbulence without pressure, developed
for the potential velocity field (the Burgers model),
does not work either since the 3D  turbulent flow
generates vorticity. In the next section we
present a theory of supersonic turbulence that explains the
obtained spectra
on the basis of the so-called She--L\'ev\^eque model of strong
turbulence. This explanation was suggested in \citep{Boldyrev},
and was motivated by recent successful
application of the
model to {\em incompressible} MHD turbulence \citep{Biskamp,Muller}.

\section{Kolmogorov--Burgers model for supersonic turbulence}
\label{theory}

The analytical model hinges on the observation
that in the inertial range the turbulence is mostly incompressible,
obeying the Kolmogorov naive scaling laws of velocity fluctuations
(see below), while in the dissipative range it behaves
as Burgers turbulence, developing shock singularities. A model
that relates the dissipative structures to the velocity scaling in the
inertial interval was developed by \citet{She-Leveque1} and \citet{She-Leveque2}.
As was pointed out by \citet{Dubrulle}, this model
represents the velocity energy cascade as a $\log$-Poisson process
and can be obtained as a suitable limit of the so-called
random $\beta$~model of turbulence \citep[see, e.g.,][]{Frisch2}. In general,
the She and L\'ev\^eque model has three input parameters.
Two of them are the exponents of the
naive (i.e. non-intermittent) scalings of the
velocity fluctuation,~$u_l\sim l^{\Theta}$, and of
the characteristic time of the energy transfer at this
scale,~$t_l\sim l/u_l\sim l^{\Delta}$.
In our case these parameters are related,~$\Delta=1-\Theta$, although
in general this does not need to be so, as e.g. in the
Iroshnikov-Kraichnan model of incompressible MHD
turbulence \citep{Grauer,Politano}.
The other parameter is the dimension of the most singular
dissipative structure,~$D$. The She--L\'ev\^eque formula, as generalized
by \citet{Dubrulle}, then
reads:
\begin{eqnarray}
\zeta(p)/\zeta(3)=\Theta \left(1-\Delta\right)p +(3-D)
\left( 1- \Sigma^{\Theta p} \right),
\label{structure_functions_general}
\end{eqnarray}
where~$\Sigma=1-\Delta/(3-D)$.
Since in our case the velocity field in the inertial
interval is mostly incompressible, one can imagine that
the energy transfer  is
due to the Kolmogorov cascade, i.e., $\Theta=1/3$, $\Delta=1-\Theta=2/3$,
while
the dissipative structures are shocks, not filaments, so~$D=2$. With
such input parameters, the formula gives:
\begin{eqnarray}
\zeta(p)/\zeta(3)=\frac{p}{9} +1 - \left(\frac{1}{3}\right)^{p/3},
\label{structure_functions}
\end{eqnarray}
For the first five structure functions the model gives:
$\zeta(1)/\zeta(3)=0.42$,
$\zeta(2)/\zeta(3)=0.74$, $\zeta(4)/\zeta(3)=1.21$,
$\zeta(5)/\zeta(3)=1.40$, in excellent agreement with
the numerical results (cf.\ Fig.~\ref{structure}). To find the absolute scalings, we need
to know the scaling of at least one of the structure functions.
As we mentioned in Sec.~\ref{numerics}, the Reynolds number is
not large enough to see good scaling of the structure functions.
However, it is known that the Fourier transform of
the second order structure function, i.e., the velocity spectrum,
can have larger scaling interval than the second order
structure function itself. Thus, knowing from the measurement of the
spectrum that~$\zeta(2)\simeq 0.7\cdots 0.8$,
we infer~$\zeta(3)\simeq 0.95\cdots 1.08$,
and the Kolmogorov relation, $\zeta(3)=1$, may indeed
hold in the inertial interval.

Note that our
formula~(\ref{structure_functions})  coincides
with the formula derived in \citep{Biskamp,Muller} for
incompressible MHD turbulence, where~$\zeta(3)=1$, and in this sense
both systems belong to the same {\em class of universality},
in accord with the ideas put forward by \cite{Dubrulle} and
\cite{She-Leveque2}. However, these two systems are
completely different, and the Kolmogorov relation,~$\zeta(3)=1$, that
is exactly proved in MHD, is {\em not} rigorously established
in our case, but rather inferred from the numerical
simulations. Strictly speaking, the scaling exponents can
be different for these systems; it is their ratios that are
reliably obtained from our numerical simulations and are believed
to be universal. For instance, there is a possibility that these
exponents can vary from cloud to cloud, depending on
the equation of state, mechanisms of dissipation, Mach numbers, etc.

For completeness, we would like to present here a cartoon model
which leads to the She-L\'ev\^eque formula. Our discussion is rather similar to
the work of \citet{Dubrulle}. The reader not interested in the
details
of the derivation can safely skip to  Sec~\ref{conclusions}.
We start with the simple $\beta$~model of turbulence, formulated
in a way suitable for
generalization of the results to the random $\beta$~model (a detailed
discussion of such models may be found in \citep{Frisch2}).
Let us assume that our system is divided into boxes of size~$l$ and
let us concentrate on the density
of the velocity energy flux over scales in each
box,~$\epsilon_l$. Now let us divide each box
into smaller boxes, with size~$l\Gamma$, where~$\Gamma <1$. The number
of boxes increases, but now we want to leave only a fraction~$\beta$
of all newly formed boxes, and since the energy flux is
constant,~$\langle \epsilon_l \rangle=const$, the density of the
energy flux in each box should be increased by~$1/\beta$. After
that, start decreasing the size of boxes again and repeat all
the procedure exactly as in the previous step. One may
introduce~$W=\epsilon_{i+1}/\epsilon_i$, which
takes the value~$1/\beta$ with probability~$\beta$,
and~$0$ with probability~$1-\beta$ at each step,
independently of previous steps. The fraction of space
occupied by boxes decreases
at each step, and one can easily check that eventually our boxes
will cover the structure with fractal
dimension
\begin{eqnarray}
D=3-\log (\beta) /\log(\Gamma).
\label{dimension}
\end{eqnarray}
This is the dimension of the structure  to which the cascade converges.
It is easy to see
that~$\langle \epsilon_l^p \rangle \sim l^{\tau(p)}$, where
\begin{eqnarray}
\tau(p)= \log \langle W^{p} \rangle/\log \Gamma.
\label{tau}
\end{eqnarray}
Let us now apply this model to incompressible turbulence. According to
the Kolmogorov refined similarity hypothesis,~$\epsilon_l$ scales
as~$u_l^3/l$; we will denote this as~$\epsilon_l \approx  u_l^3/l$.
So we have
\begin{eqnarray}
\zeta(p)=p/3+\tau(p/3)=p/3+(3-D)(1-p/3),
\label{beta_model}
\end{eqnarray}
which gives, e.g.,
a velocity spectrum  steeper than the Kolmogorov one.
The $\beta$ model produces the {\em linear} relation~(\ref{beta_model})
which is not what is observed.

We now generalize this model in the
following way. First, let us assume that the steps of size changes
are very small, i.e., $\Gamma=1-x/(3-D)$, and~$\beta=1-x$,
where~$x\to 0$.
These expressions are chosen to satisfy~(\ref{dimension}) up to
the first order in small~$x$. To modify
the previous model we now assume that we do not disregard any newly
formed boxes, whose fraction would be $1-\beta=x$, but instead fill them
with~$W=\beta_1$, while the other boxes are filled with some
factor~$W=\beta_2$. This procedure leads to a so-called random
$\beta$~model. Since we have to
preserve~$\langle \epsilon_l \rangle=const$, we have to impose the
relation~${x}{\beta_1}+
{\beta_2}\left(1-x \right)=1$ which gives
$\beta_2=1+x(1-\beta_1)$. Using formula~(\ref{tau}),
we now derive~$\tau(p)=C(\beta_1-1)p+C(1-\beta_1^p)$, where
we introduced the co-dimension~$C=3-D$. If we now assume the
Kolmogorov relation~$\epsilon_l \approx  u_l^3/l$, we immediately
recover formula~(\ref{structure_functions_general}),
where~$\zeta(3)=1$
and $\beta_1$ plays the role of~$\Sigma$. This shows that~$\Sigma$
describes the degree of intermittency in the She--L\'ev\^eque model.
However, in the
compressible case the
Kolmogorov relation is not proven. We can, however, hope
that if we change the length variable~$l\to {\tilde l}= \xi(l)$ in such
a way that the scaling exponent of the third-order structure function
with respect to~$\tilde l$ becomes~${\tilde \zeta}(3)=1$,
the scaling of all the other structure functions will be restored as well.
To achieve that we have to define the new scaling variable
as~$ {\tilde l}=\xi(l)=\langle u^3_l \rangle/\langle \epsilon_l \rangle$.
Also, instead of the Kolmogorov relation,
we assume~$\epsilon_l/\langle \epsilon \rangle\propto u_l^3/\langle u_l^3 \rangle$,
which is suggested by the quadratic
nonlinearity of the Navier-Stokes equation.
Redefining
the scaling~$\tau(p)$ as $\langle \epsilon_l^p \rangle \sim {\tilde l}^{\tau(p)}$,
we now easily get:
\begin{eqnarray}
\zeta(p)/\zeta(3)=p/3-C(1-\beta_1)p/3+C(1-\beta_1^{p/3}),
\label{random_beta}
\end{eqnarray}
which is the required generalization of the She-L\'ev\^eque formula.
The unknown
parameters~$\beta_1$ and~$C$ can be obtained, e.g.,
from numerical simulations. In this formula it is not obvious
that $C$~is a co-dimension of the most singular dissipative structure,
since the cascade was organized not with respect to physical
length~$l$, but with respect to~$\tilde l$.

\section{Application to molecular clouds}
\label{conclusions}

We now ask how adequate our numerical and analytical
results are to real molecular clouds. The density structures
of molecular clouds, as obtained from observations, are rather
complicated, and one may question the validity of our simple model
assuming shock singularities. In fact, observational relation of
molecular cloud mass to the cloud size is close to~$M\sim L^{D}$,
where $D$~is estimated to be greater than~$2$ see,
e.g. \citet{Larson2}, and also \citet{Elmegreen3}, where the
estimate~$D=2.3\pm 0.3$ was presented.
If the density were concentrated in elongated shocks, the scaling
power would be 2. A plausible resolution lies in the
assumption
that over a range of scales (not available in numerical simulations
due to limited resolution) dense structures are folded to form
a complex distribution having {\em fractal} dimension~$D\geq 2$; a
complicated fractal (even multifractal) density distribution is
indeed inferred from observations \citep{Chappell}.
It seems natural to assume that the energy dissipation occurs
mostly in shocks \citep{Stone,Ostriker}, and therefore
the dissipative singularities should
have the same dimension.
Formula~(\ref{structure_functions_general}) thus relates the fractal
dimension of a given molecular cloud to scaling properties
of the velocity field in this cloud. In principle, this dimension
can be different for different clouds, depending on mechanisms of
cooling, dissipation, structure of magnetic fields, etc.
Substitution of the fiducial value~$D=2.3$ in our
formula~(\ref{structure_functions_general}) leads to even better
agreement of the theory with recent observations, see \cite{Brunt},
producing~$\langle |\Delta u| \rangle \sim L^{0.55}$
and~$E_k\sim k^{-1.83}$, where we assumed that the Kolmogorov
relation~$\zeta(3)=1$ holds.

It is interesting to note that formula~(\ref{structure_functions_general})
also gives the upper boundary for the dimension of density distribution.
The parameter~$\Sigma$ introduced in~(\ref{structure_functions_general})
measures the degree of intermittency of turbulence. For example,
for~$\Sigma\to 1$ one recovers the Kolmogorov, non-intermittent,
scaling~$\zeta(p)=p/3$. By its definition,~$0\leq \Sigma \leq 1$, and
therefore,~$D\leq 2\frac{1}{3}$. Taking into account
that on small scales the dissipative structures take the form
of two-dimensional shocks,
one may argue that the lower boundary for the density dimension
is~$D=2$. We
thus established a rather narrow window for~$D$
consistent with our theory. With the aid of
formula~(\ref{structure_functions_general}) one then
finds the corresponding
intervals for the structure functions scaling exponents that are admissible
in our theory. In particular,
the first order exponent is expected to lie in the
interval~$[0.42\cdots 0.78]$,
and the second order exponent, related to the velocity spectrum,
within~$[0.74\cdots 0.89]$. It is interesting that the observed
value, $D=2.3$,
is rather close to the upper boundary. In
Figs.\ \ref{zeta1} and \ref{zeta2} we
plotted the scaling exponents~$\zeta(1)$ and~$\zeta(2)$ as functions
of~$D$ for~$2\leq D \leq 2\frac{1}{3}$. We see that the exponents are very
sensitive to the value of~$D$, and it is therefore very  hard to
infer~$\zeta$'s from measurements of~$D$ -- this
would require measuring~$D$ with high precision. More practical
test of the theory would be plotting observed~$\zeta(2)$ versus
$\zeta(1)$, our theory predicts that the values should lie on
the line shown in Fig.~\ref{ratios}.

\section{Conclusions}
\label{conclusions1}
The presented model proves to be rather successful in explaining the
statistical properties of supersonic turbulence. In particular,
we were able to explain the observational scaling
relations of the turbulence in molecular clouds.
It would be highly
desirable to obtain a clear physical picture underlying
the Log-Poisson energy cascade assumed in the
derivation of~(\ref{structure_functions_general}). Also, the
relation of the density distribution to the statistics of the velocity
field requires a better understanding.
Such a relation may be derived analytically in a simplified
one-dimensional case \citep{Boldyrev1,Boldyrev-Brandenburg}, although in the
three-dimensional case the question is still open. However,  in the derivation of
formula~(\ref{random_beta}) we explicitly constructed the {\em multifractal}
distribution of the energy dissipation~$\epsilon$, which can possibly
be related to the
multifractal distribution of the density, since the energy dissipates in shocks where
the density is accumulated.  And finally, in the present paper we did not try to
address the problem of the initial mass distribution of stars \citep[but see][]{PN}.
The process of
turbulent density fragmentation leads to creation of complicated density
structures and sets the {\em initial conditions} for gravitationally unstable
density clumps. The effects of self-gravity, and the dynamics of collapsing density
cores fall beyond the scope of the present paper. All these are the questions for the
future.


\newpage

{
\begin{figure} [tbp]
\centerline{\psfig{file=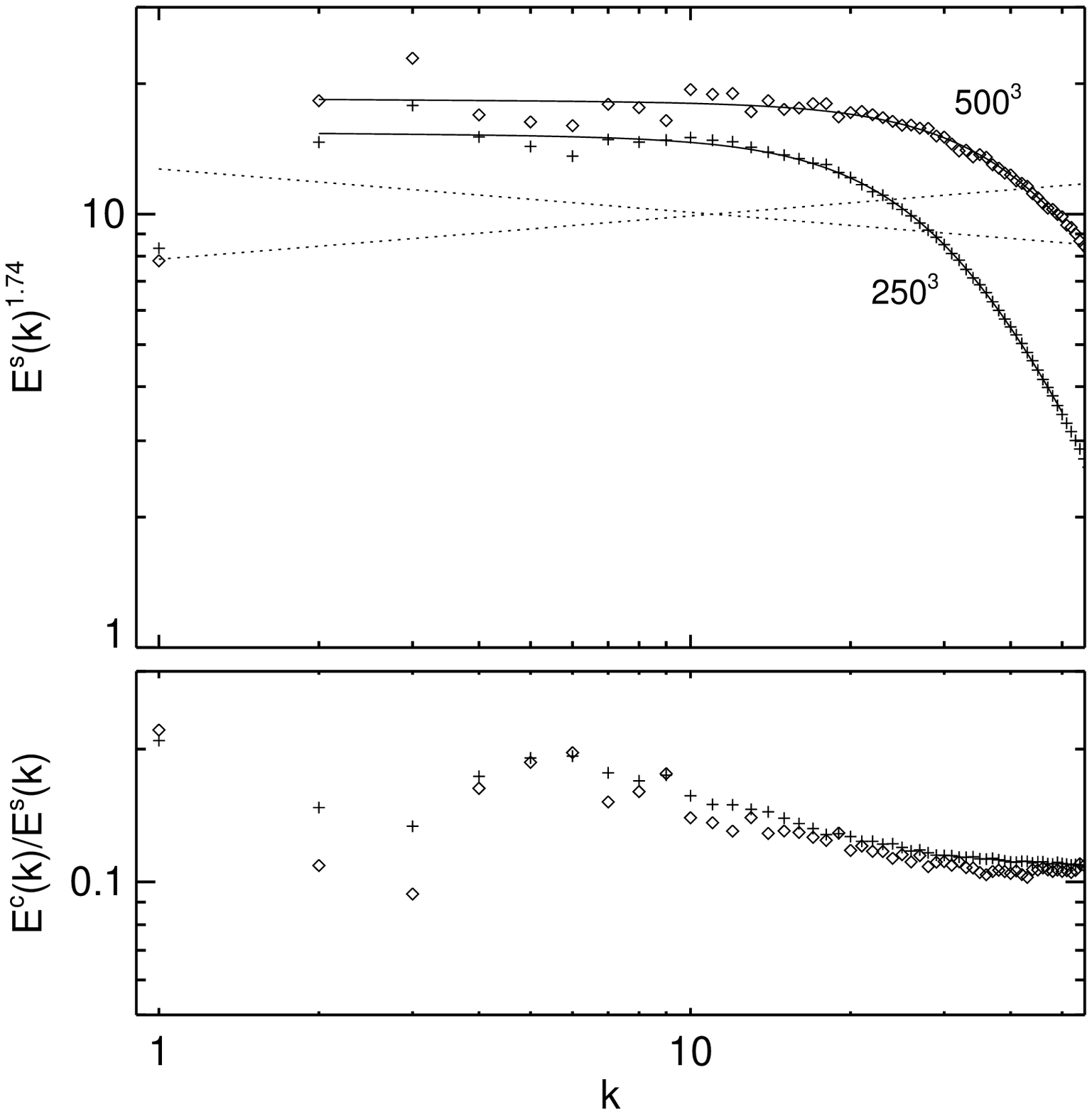,width=6in}}
\caption{Power spectra of driven, super--sonic turbulence.
The top panel shows the solenoidal power, compensated by $k^{1.74}$,
and averaged over two turn-over times, after about three t.o. times from
the start of the run, in a numerical experiment with resolution $250^3$
(plusses), using random driving at $1\leq k\leq2$.
Additional data points (diamonds) are from a $500^3$ experiment, continued for
about two tenths of a turn-over time from a $250^3$ snap shot, which is
enough to establish the extended large $k$ solenoidal spectrum.
The solid lines are least squares fits of the data in the range
$2 \leq k \leq 50$ to functional forms $a / (k^p + b k^q)$, where
$p$ is the power law index in the inertial range (1.75 for the
$250^3$ case and 1.74 for the $500^3$ case).
Dashed lines show comparison slopes with spectral indices
different by $\pm 0.1$, from~$1.74$. The bottom panel shows the average
ratio of compressional to solenoidal power in the same experiments.}
\label{spectra}
\end{figure}
}


{
\begin{figure} [tbp]
\centerline{\psfig{file=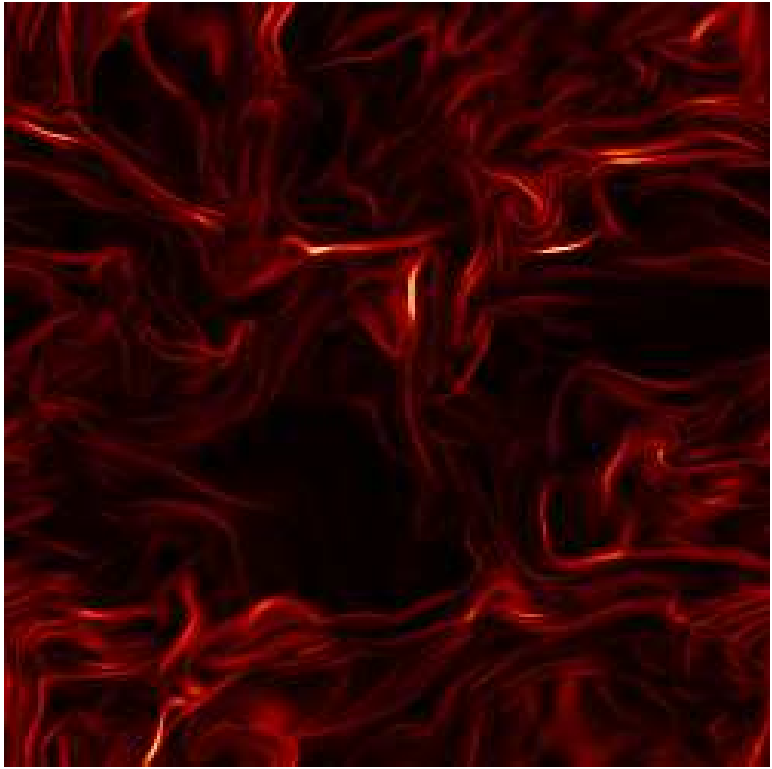,width=4in}}
\vskip5mm
\caption{A random, two-dimensional cut through the physical
simulation domain of a snapshot from a $250^3$ experiment. The
cross-section of the density field shows filamentary structures
that correspond to sheet-like shock density structures.}
\label{shocks}
\end{figure}
}

{
\begin{figure} [tbp]
\centerline{\psfig{file=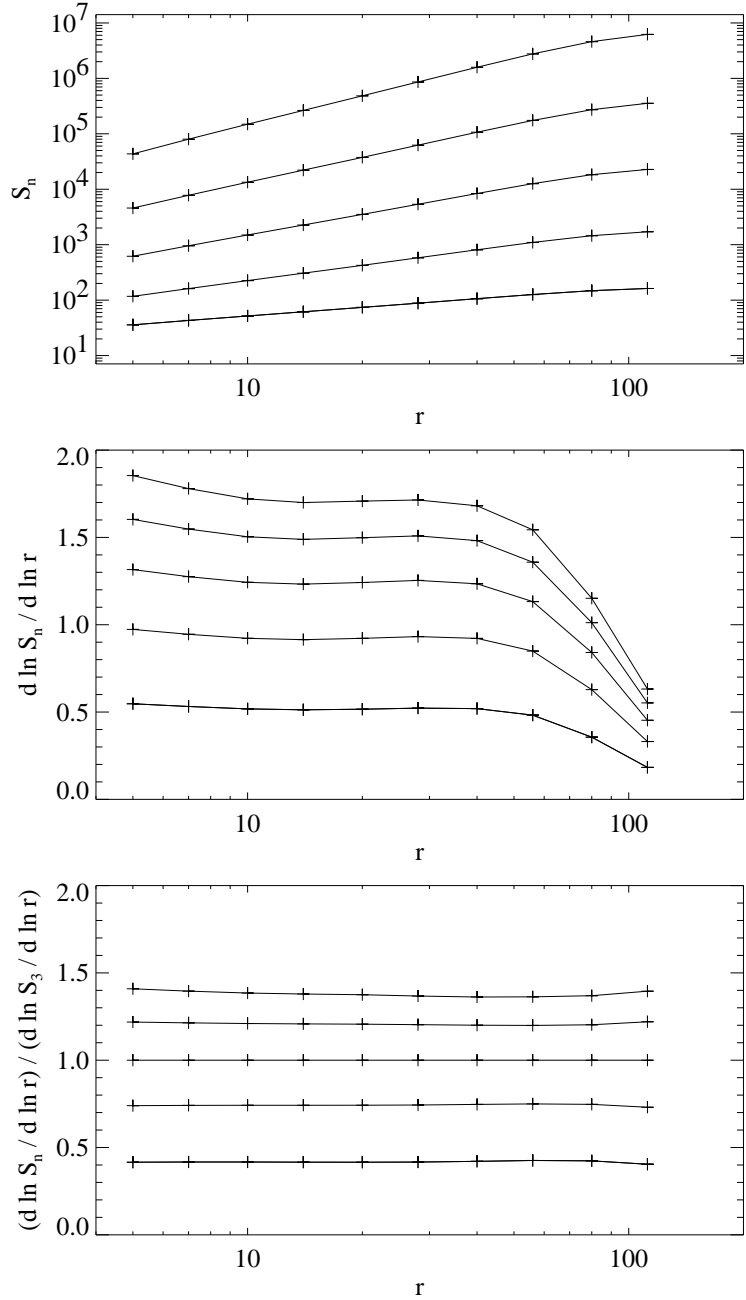,width=4.in}}
\caption{Transversal structure functions computed for~$p=1,2,3,4,5$
(correspondingly, from bottom to top in each panel). The first
panel shows Log-Log
plots of the structure functions.  The second panel shows the differential
slopes,~$\zeta(1),\dots , \zeta(5)$. The scaling range
is very short, due to the limited Reynolds number.
The third panel presents the ratios of the differential slopes
to~$\zeta(3)$, which exhibit excellent scalings, in agreement with the
Extended Self-Similarity hypothesis. These ratios are well
described by our formula~(\ref{structure_functions}).}
\label{structure}
\end{figure}
}

{
\begin{figure} [tbp]
\centerline{\psfig{file=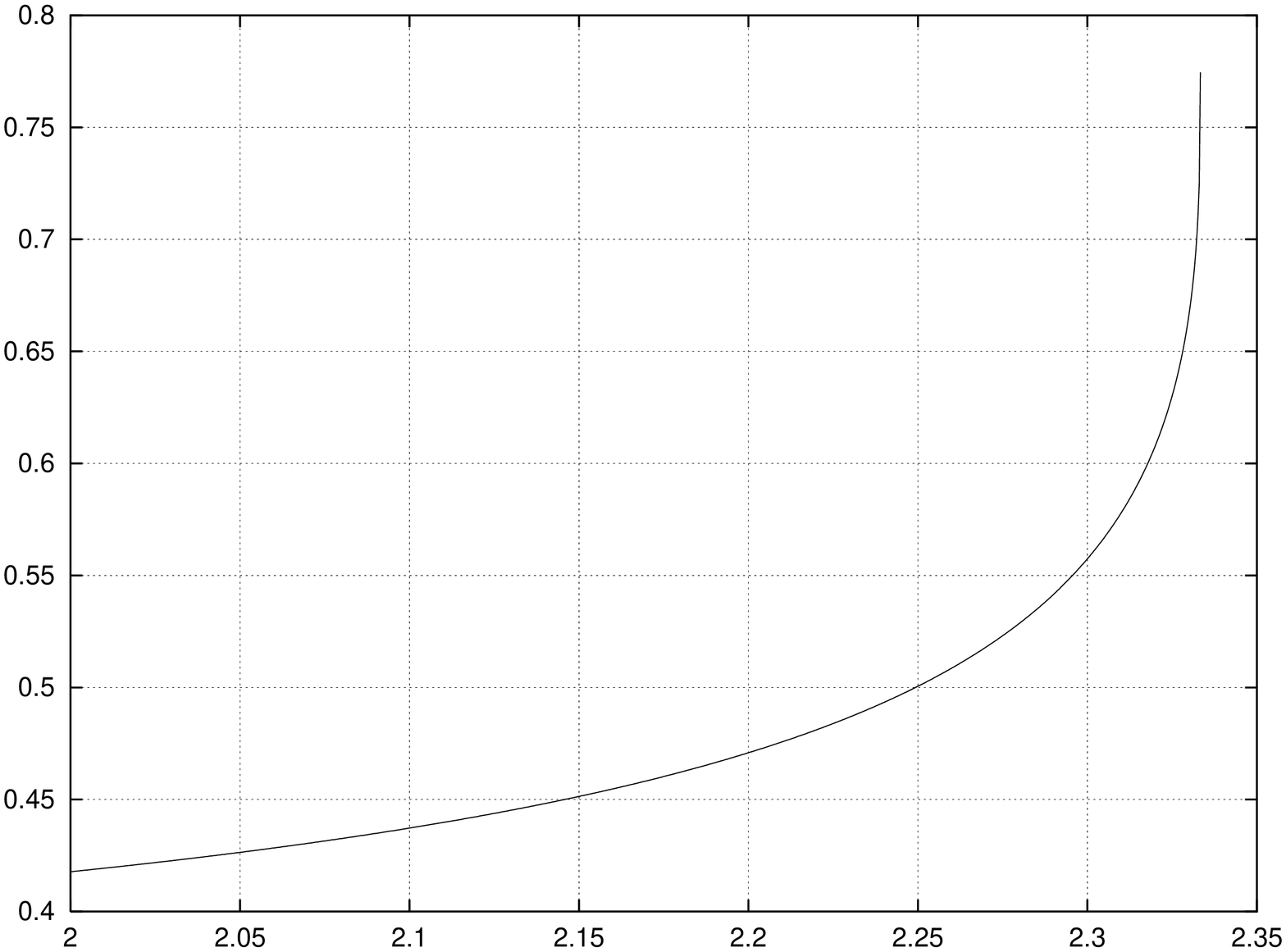,width=5.5in,angle=-90}}
\caption{Normalized scaling exponent of
the first-order structure
function,~$\zeta(1)/\zeta(3)$, as a function of the
dimensionality,~$D$, of the most
singular dissipative structure, which can be close to the fractal
dimensionality of the cloud. We use
formula~(\ref{structure_functions_general})
with~$\Theta=1/3$, $\Delta=2/3$ and $D$~changes from~$2$ to~$2\frac{1}{3}$.
The results of our numerical simulations correspond to~$D=2$.}
\label{zeta1}
\end{figure}
}
{
\begin{figure} [tbp]
\centerline{\psfig{file=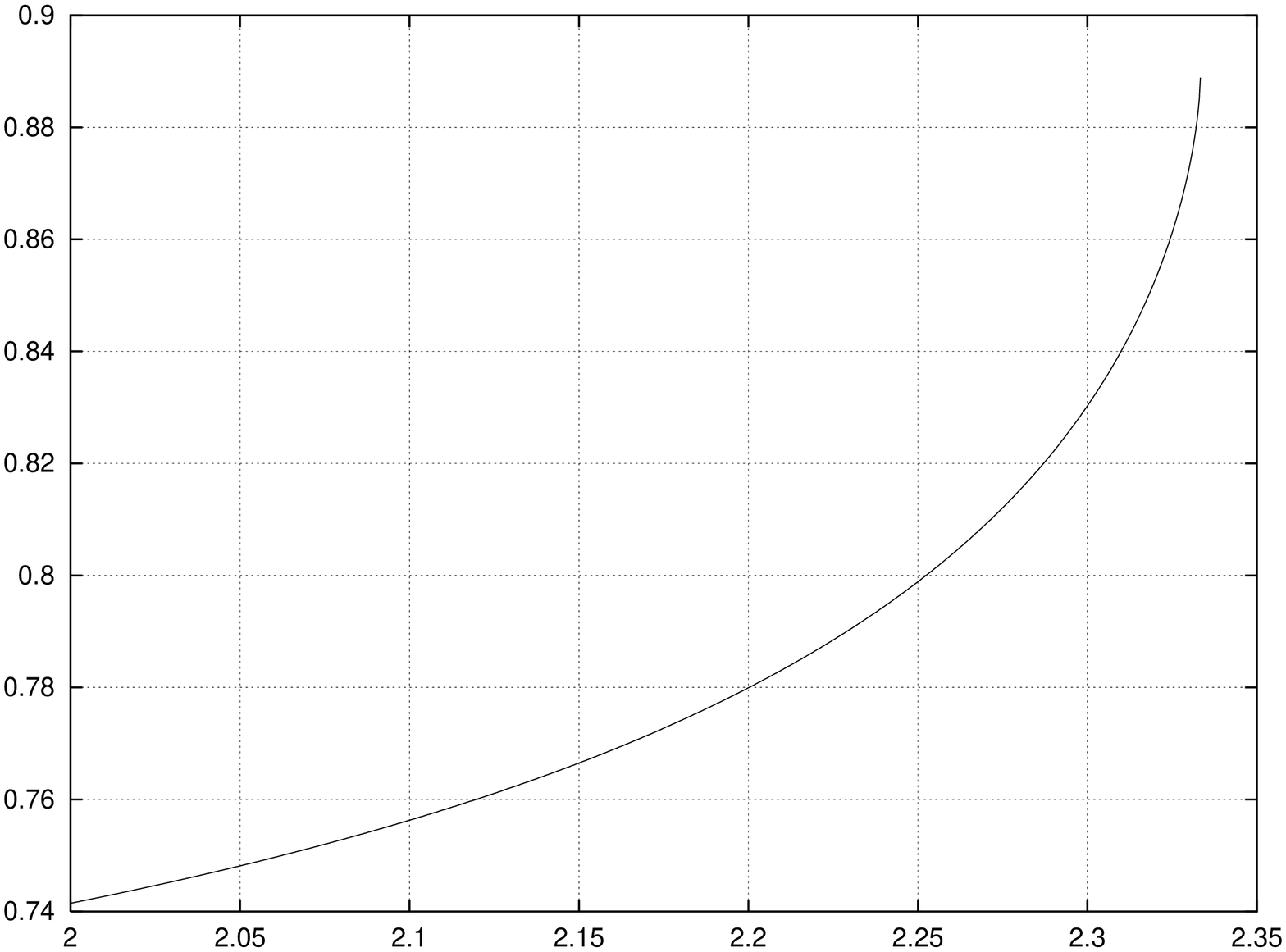,width=5.5in,angle=-90}}
\caption{Normalized scaling exponent of the
second-order structure
function,~$\zeta(2)/\zeta(3)$, as a function of the
dimensionality,~$D$, of the most
singular dissipative structure, which can be close to the fractal
dimensionality of the cloud. We use
formula~(\ref{structure_functions_general})
with~$\Theta=1/3$, $\Delta=2/3$ and $D$~changes from~$2$ to~$2\frac{1}{3}$.
The results of our numerical simulations correspond to~$D=2$. }
\label{zeta2}
\end{figure}
}

{
\begin{figure} [tbp]
\centerline{\psfig{file=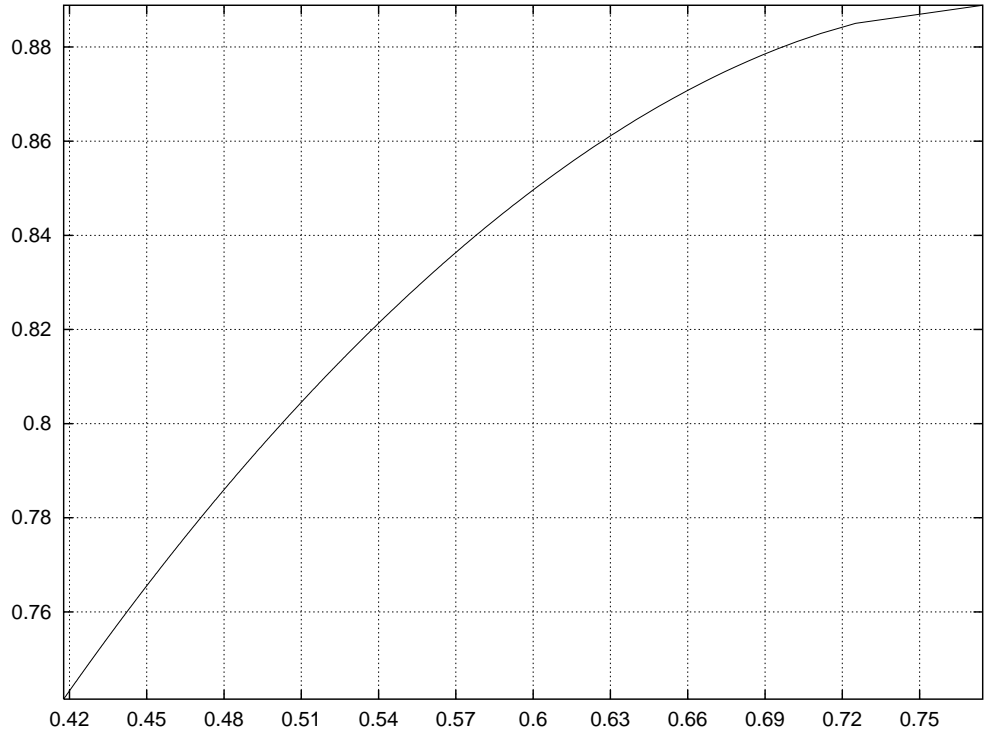,width=5.5in,angle=-90}}
\caption{Normalized scaling exponent~$\zeta(2)/\zeta(3)$ (vertical axis)
as a function of the normalized scaling
exponent~$\zeta(1)/\zeta(3)$ (horizontal axis), as given by
formula~(\ref{structure_functions_general})
with~$\Theta=1/3$ and~$\Delta=2/3$. The corresponding change of
parameter~$D$ is in the interval~$[2;2\frac{1}{3}]$. The results of
our numerical simulations correspond to~$D=2$,
i.e.~$\zeta(1)/\zeta(3)=0.42$ and~$\zeta(2)/\zeta(3)=0.74$.}
\label{ratios}
\end{figure}
}

\end {document}